\documentclass[11pt,a4paper]{article}
\pdfoutput=1
\usepackage{jheppub_kim}

\usepackage{multirow, graphicx,amssymb,url,mathrsfs,amsmath}
\usepackage{wrapfig,boxedminipage,setspace,subfigure,epsfig}
\usepackage{amsxtra,amstext,latexsym,dsfont,amsfonts}
\usepackage{color,eucal}

\newcommand{\be}{\begin{equation}}
\newcommand{\ee}{\end{equation}}

\newcommand{\dd}{\mathrm{d}}

\title{Gauge Invariance and Holographic Renormalization}

\author[a]{Keun-Young Kim,}
\author[a]{Kyung Kiu Kim,}
\author[b]{Yunseok Seo,}
\author[c,d]{and Sang-Jin Sin}

\emailAdd{fortoe@gist.ac.kr}
\emailAdd{kimkyungkiu@gmail.com}
\emailAdd{yseo@hanyang.ac.kr}
\emailAdd{sjsin@hanyang.ac.kr}

\affiliation[a]{ School of Physics and Chemistry, Gwangju Institute of Science and Technology,
Gwangju 500-712, Korea
}
\affiliation[b]{Research Institute for Natural Science, Hanyang University, Seoul 133-791, Korea}
\affiliation[c]{ Department of Physics, Hanyang University, Seoul 133-791, Korea }
\affiliation[d]{ School of Physics, Korea Institute for Advanced Study, Seoul 130-722, Korea }

\abstract{We study the gauge  invariance of physical observables in holographic theories
under the local diffeomorphism. We find that gauge invariance is intimately related to the holographic renormalisation: the local counter terms defined in the boundary cancel most of gauge dependences of the on-shell action as well as the divergences.
There is a mismatch in the degrees of freedom  between the bulk theory and   the boundary one.
We resolve this problem  by noticing that there is a residual gauge symmetry(RGS).
By extending the RGS such that it satisfies infalling boundary condition at the horizon,  we can understand
the problem in the context of general holographic embedding of a   global symmetry at the boundary  into the local gauge symmetry in the bulk.}


\subheader{ $ $}

\arxivnumber{1502.02100}

\begin{document}
\maketitle
\section{Introduction}
According to AdS/CFT correspondence, any global symmetry at the boundary theory is lifted to a local symmetry
in the bulk~\cite{Witten:1998qj, Hartnoll:2009sz}. The gauge symmetry is essential to reduce the degree of freedom which is enlarged by going into one higher dimension.  The physical goal in holography is the boundary quantities which does not know the presence of higher dimension or gauge degrees of freedom, while we use the tools  in the bulk theory. Therefore the   gauge invariance of a physical quantity is a critical issue for the validity of the AdS/CFT. Also tracing the gauge invariance gives much intuition on the way how holography actually works, especially how global symmetry is encoded in the local gauge symmetry.

One can find gauge invariant combinations of the fields,  and express the physical quantities in terms of such master variables, however, it is not always  easy to find such gauge invariant combination.
Even in the case they are available, it is not very convenient
to use such fields, especially  if many fields are coupled, because the physical quantities are defined in terms of the field variables which are formally gauge dependent.
For example \cite{Hartnoll:2009sz}, energy momentum tensor and chemical potential
is defined  in terms of metric/gauge field  which are not gauge invariant. Similarly, heat currents  can be related to the metric perturbation
defined only in a specific gauge where time period has definite relation with temperature.

In recent works \cite{Kim:2014bza,Kim:2015dna}, based on \cite{Amado:2009ts,Kaminski:2009dh}, we developed a systematic method to numerically calculate the Green's functions and all AC transports quantities simultaneously for the case where many fields are coupled and there are {\it constraints} due to gauge symmetry. Although we have tested the validity of the  procedure by showing the agreement of
 zero frequency limits of AC conductivities with the known analytic DC conductivities \cite{Andrade:2013gsa,Donos:2014cya,Donos:2014uba}
 we still think that we need to prove the gauge invariance of our procedure as a matter of principle.
We  found that the   bulk gauge invariance is intimately related to the holographic renormalisation. Although the local counter terms were introduced to kill the divergences, it also kills most of gauge dependence.

Furthermore, there is a  residual gauge symmetry(RGS) even after we fix the axial gauge $g_{rx}=0$.  While equations of motion can be written  in terms of the
the gauge invariant master fields ${\cal P}_{h}, {\cal P}_{\chi}$ \eqref{ginvp},
it turns out that the quadratic on-shell action, the generating function for two point retarded Green's functions, can not be written as such. However, we prove that the Green's functions are still invariant under such a symmetry.

There is a mismatch in the degrees of freedom in the bulk and
those at the boundary: there are only two independent bulk solutions satisfying the in-falling boundary conditions  while we need three  solutions at the boundary since there are three independent  source fields.
The RGS is the one that resolves the problem:
since it cannot satisfy a proper boundary condition,  it is not
a proper gauge symmetry but a `solution generating symmetry'.
It generate the desired solution at the boundary and therefore we should accept its bulk counter part as a new physical degree of freedom as well although   it can not satisfy the infalling boundary condition (BC).
By extending the RGS such that it satisfies infalling boundary condition at the horizon,  we can make the bulk solution more natural in the sense that it satisfies the infalling BC.
With  such solution we can also understand the problem in the context of general structure of holography, namely the correspondence between a   global symmetry at the boundary and the local gauge symmetry in the bulk.

\section{Action and background solution}

Let us first briefly review the system we will discuss, which has been analysed in detail in \cite{Andrade:2013gsa,Kim:2014bza,Bianchi:2001kw}.
The holographically renormalised action($S_\mathrm{ren}$) is given by
\begin{equation} \label{SS}
S_\mathrm{ren} = S_{\mathrm{EM}}  + S_\psi + S_\mathrm{c} \,,
\end{equation}
where
\begin{equation}
S_{\mathrm{EM}} = \int_{M}  \dd^{4}x \sqrt{-g} \left[   R - 2 \Lambda -\frac{1}{4}F^2    \right]
- 2 \int_{\partial M} \dd^3 x   \sqrt{-\gamma} K \,,
\end{equation}
is the usual action for charged black hole in AdS space($\Lambda < 0$) with the Gibbons-Hawking term and
\begin{equation}
S_\psi = \int_M \dd^{4}x \sqrt{-g} \left[  - \frac{1}{2}\sum_{I=1}^{2} (\partial\psi_I)^2  \right] \, ,
\end{equation}
 is the action for  two free massless scalars added for a momentum relaxation effect.
$S_c$ is the counter term
\begin{align} \label{Sc}
S_\mathrm{c}= \eta_c \int_{\partial  M}  \dd x^3 \sqrt{-\gamma} \left( - 4 -R[ \gamma]  + \frac{1}{2}  \sum_{I=1}^2 \gamma^{\mu\nu}  \partial_\mu \psi_I \partial_\nu\psi_I    \right) \,,
\end{align}
which is included to cancel   the divergence in
$S_{\mathrm{EM}} + S_\psi $.
 Here we introduced $\eta_c$ to keep track of the effect of the counter term. At the end of the computation we will set $\eta_c = 1$.

The action \eqref{SS} yields general equations of motion\footnote{Index convention: $M,N,\cdots = 0,1,2,r$, and $\mu,\nu,\cdots = 0,1,2$, and $i,j,\cdots = 1,2$.}
\begin{align}
&R_{MN} = \frac{1}{2}g_{MN} \left( R - 2 \Lambda -\frac{1}{4}F^2  - \frac{1}{2}\sum_{I=1}^{2} (\partial\psi_I)^2 \right) +\frac{1}{2}  \sum_{I}  \partial_M \psi_I \partial_N \psi_I +\frac{1}{2} {F_M}^P F_{NP} \,,
\label{Rmn} \\
&\nabla_M F^{MN} =0 \,,  \qquad
\nabla^2 \psi_I =0 \,, \label{Epsi}
\end{align}
which admit the following solutions
\begin{align}
\dd s^2 &= G_{MN} \dd x^{M} \dd x^{N} =  -  f(r) \dd t^2 +  \frac{\dd r^2}{f(r)}  +  r^2 \delta_{ij} \dd x^i \dd x^j \,,  \label{groundsol1}  \\
& \quad f(r)= r^2 - \frac{ \beta^2}{ 2 } - \frac{m_0}{r} + \frac{ \mu^2   }{4} \frac{r_0^{2}}{r^{2}} \,, \qquad
m_0 = r_0^3 \left(  1+\frac{\mu^2}{4 r_0^2} - \frac{\beta^2}{2 r_0^2}     \right) \,,  \label{groundsol2} \\
A&= \mu \left(  1- \frac{r_0}{r}   \right)\dd t    \,,  \label{groundsol3} \\
\psi_I &= \beta_{Ii} x^i = \beta \delta_{Ii} x^i\,. \label{groundsol4}
\end{align}
These are reduced to AdS-Reissner-Nordstrom(AdS-RN) black brane solutions when $\beta=0$. Here we have taken special $\beta_{Ii}$, which satisfies $\frac{1}{2}\sum_{I=1}^2   \vec\beta_{I} \cdot \vec\beta_{I}= \beta^2 $ for general cases.

The solutions \eqref{groundsol1} - \eqref{groundsol4} are characterised by three parameters: $r_0$, $\mu$,  and $\beta$.  $r_0$ is the black brane horizon position($f(r_0) =0$)
and can be replaced by temperature $T$ for the dual field theory:
\begin{equation}
T = \frac{f'(r_0)}{4\pi} = \frac{1}{4\pi} \left( 3r_0 - \frac{\mu^2+ 2\beta^2}{4r_0}  \right) \, .
\end{equation}
Non-vanishing components of energy-momentum tensor and charge density read
\begin{equation} \label{Ttt}
\langle{T^{tt}}\rangle = 2m_0 \,, \quad \langle{T^{xx}}\rangle = \langle{T^{yy}}\rangle = m_0 \,, \quad
\langle J^t \rangle = \mu r_0 \,.
\end{equation}
{$\langle{T^{tt}}\rangle  = 2 \langle{T^{xx}}\rangle$ implies that charge carriers are still of massless character.} From here we set $r_0 = 1$ not to clutter.

\section{Gauge fixing and residual gauge transformation }

To study electric, thermoelectric, and thermal conductivities we introduce  small fluctuations around the background \eqref{groundsol1} - \eqref{groundsol4}
\begin{align}
 \delta A_x(t,r) &= \int^{\infty}_{-\infty} \frac{\dd \omega}{2\pi}  e^{-i\omega t}  a_{x}(\omega,r) \,,   \label{fluc1} \\
\delta g_{tx}(t,r) &=  \int^{\infty}_{-\infty} \frac{\dd \omega}{2\pi} e^{-i\omega t} r^2 h_{tx}(\omega,r)\,,  \label{flucGtx} \\
\delta g_{rx}(t,r) &=  \int^{\infty}_{-\infty} \frac{\dd \omega}{2\pi} e^{-i\omega t} r^2 h_{rx}(\omega,r)\,,  \label{flucGrx} \\
 \delta \psi_1(t,r) &= \int^{\infty}_{-\infty} \frac{\dd \omega}{2\pi} e^{-i\omega t}  \chi (\omega,r) \,. \label{fluc4}
\end{align}
The fluctuations are chosen to be independent of $x$ and $y$.
This is allowed since all the background fields appearing in the equations of motion turn out to be independent of $x$ and $y$.
The gauge field fluctuation($\delta A_x(t,r)$) sources metric($\delta  g_{tx}(t,r), \delta  g_{rx}(t,r) $) and scalar field($\delta \psi_1(t,r)$) fluctuation and vice versa and all the other fluctuations are decoupled.
We will work in momentum space and $h_{tx}(\omega, r)$ and $h_{rx}(\omega, r)$ is defined so that it goes to constant as $r$ goes to infinity.



By linearising the full equation of motion, we  get four equations. However one of them can be obtained by the others. Thus we may consider following three equations:
\begin{align}
&(\chi ' -\beta  h_{rx} )-\frac{i \mu  \omega  a_x }{\beta  r^2 f(r)} -\frac{i r^2 \omega (h_{tx}' +i \omega h_{rx}  )}{\beta  f(r)}=0\,, \\
&a_x''(r)+\frac{a_x'(r) f'(r)}{f(r)}+\frac{\omega ^2 a_x(r)}{f(r)^2} +\frac{\mu (h_{tx}' +i \omega h_{rx}  )}{f(r)}=0 \,, \\
& f(r)f'(r) (\chi '(r)-\beta  h_{rx})+ f(r)^2(\chi ' -\beta  h_{rx})' +\frac{2 f(r)^2(\chi ' -\beta  h_{rx} ) }{r}  \nonumber \\
& \qquad \qquad \qquad \qquad \qquad \qquad \qquad \qquad \qquad \quad + \omega ^2 \chi (r)-i \beta  \omega h_{tx}(r)  =0 \,.
\end{align}
If we differentiate  the third equation with respect to $r$, all equations can be written in terms of three variables, $\mathcal{P}_\chi,\mathcal{P}_h$, and $a_x$, where
\begin{align} \label{ginvp}
\mathcal{P}_\chi \equiv \chi ' -\beta  h_{rx}\,, \qquad \mathcal{P}_h \equiv h_{tx}' +i \omega h_{rx}  \,.
\end{align}
Therefore, $h_{rx}$ is a non-dynamical degree of freedom. Indeed, $\mathcal{P}_\chi,\mathcal{P}_h$, and $a_x$ are invariant under a diffeomorphism generated by $\xi^\mu = (0,\zeta(r)e^{-i\omega t},0,0)$, under which the fields are  transformed as follows:
%
\begin{align} \label{hhh1}
&\delta h_{rx} =  \frac{1}{r^2}( \nabla_r \xi_x +  \nabla_x \xi_r ) = \zeta'(r) e^{-i\omega t} \,, \\
&\delta h_{tx} =  \frac{1}{r^2}( \nabla_t \xi_x +  \nabla_x \xi_t )= -i\omega \zeta(r) e^{-i\omega t} \,, \\
&\delta \chi =\beta \zeta(r) e^{-i\omega t}\,, \\
&\delta a_x =0 \,. \label{hhh2}
\end{align}
Using this  gauge degree of freedom, one may set $h_{rx}=0$, which is so called the axial gauge.
The numerical calculation in \cite{Kim:2014bza} has been performed in this gauge.
A question arises whether the resulting physical quantities are independent of such
gauge fixing condition.

Furthermore, even after we fix $h_{rx}=0$, one can still find a residual gauge transformation which is given by constant $\zeta$~\cite{Donos:2013eha}. This residual diffeomorphism doesn't change the gauge fixing condition $h_{rx}=0$ and generates constant shift on $h_{tx}$ and $\chi$, because the equations of motion contain only derivatives of $h_{tx}$ and $\chi$ and the linear combination of them,
$  \omega   \chi (r)-i \beta    h_{tx}(r)$, which is invariant under
\be \label{residual1}
h_{tx} \to h_{tx} +h_{0}, \hbox{ and } \chi  \to \chi + i \frac{\beta}{\omega} h_{0},
\ee
where $h_{0}$ is a constant.
Thus  there is one parameter  constant  solution given by
\be
a_{x}=0, \quad h_{tx}=h_{0}, \quad  \chi=i \frac{\beta}{\omega} h_{0} \,, \label{RGS}
\ee
which does not satisfy in-falling boundary condition so it is not a physical degree of freedom\footnote{It is a regular solution at future horizon.}.
 We call it the residual gauge symmetry(RGS) because it is generated by the zero mode of a diffeomorphism generator.
This kind of solution was first introduced in \cite{Hartnoll:2007ai}

 Why should there be such a residual degree of freedom?
 It can be traced to the difference of the differential equation near  horizon and
those near boundary.
Near the black hole horizon ($r \rightarrow 1$) the solutions are expanded as
\begin{equation} \label{nearh}
\begin{split}
&h_{tx} =  (r-1)^{\nu_\pm+1}\left(h_{tx}^{(I)} + h_{tx}^{(II)}(r-1) + \cdots \right) ,  \\
&a_x=  (r-1)^{\nu_\pm}\left(a_x^{(I)} + a_x^{(II)}(r-1) + \cdots \right)  ,\\
&\chi=  (r-1)^{\nu_\pm}\left(\chi^{(I)} + \chi^{(II)}(r-1) + \cdots \right) ,
\end{split}
\end{equation}
where $\nu_\pm = \pm i 4 \omega  /(-12  + 2\beta^2 + \mu^2)= \mp i\omega/(4\pi T)$ and the incoming boundary condition corresponds to $\nu = \nu_+$.
By inserting these to the equations of motion,
one can easily find a linear relations between the zero-th modes:
\be \label{const1}
(\nu+1) h_{tx}^{(I)}+ { \mu} a_{x}^{(I)}+  { \beta} \chi^{(I)} =0,
\ee
Notice that  all other modes are generated by these. Thus
there is a well defined  constraint equation  which reduces the degrees of freedom.

On the other hand, by inserting the expansion near the boundary ($r \rightarrow \infty$)
\begin{equation} \label{nearb}
\begin{split}
h_{tx} &=   h^{(0)}_{tx} + \frac{1}{r^2} h^{(2)}_{tx} + \frac{1}{r^3}h_{tx}^{(3)}+\cdots,  \\
a_x&=a_x^{(0)} + \frac{1}{r}a_x^{(1)}+ \cdots, \qquad  \quad \\
 \chi &= \chi^{(0)} + \frac{1}{r^2} \chi^{(2)}+ \frac{1}{r^3}\chi^{(3)} + \cdots,
\end{split}
\end{equation}
to the equations of motion, we can not get any relation between the zero-th modes
$a_x^{(0)} , h^{(0)}_{tx}$, and $\chi^{(0)}$, all of which are related to the higher modes.
More explicitly,
\begin{equation} \label{gaugecom}
 \omega (\omega\chi^{(0)}-  {i\beta} h^{(0)}_{tx} ) -2\chi^{(2)}=0,
 \quad  i\beta   (\omega\chi^{(0)}-  {i\beta}h^{(0)}_{tx} )-2 h_{tx}^{(2)}=0,
\end{equation}
which are evolution equations in $r$-direction.
Therefore, there is no constraint equation. Then there is a crisis of mismatch of degrees of freedom  and this crisis is resolved by the effective residual degree of freedom described above. However, this residual gauge degree of freedom raises
another issue of  invariance of physics under this symmetry.  We will address this issue at the end
of section \ref{sec5}.

\section{Holographic renormalization and  gauge invariance}
Now we come back to the question whether physical quantities are independent of
the choice of the gauge condition $h_{rx}(r)=0$.
We will show this by proving that the generating function of physical quantities, the on-shell action,  is invariant even in the case with $h_{rx}(r) \ne 0$.

The on-shell renormalised action to quadratic order in fluctuation fields, $S_{\mathrm{ren}}^{(2)} $, is

\begin{equation}\label{Sreal}
\begin{split}
S_{\mathrm{ren}}^{(2)}
= \lim_{r \rightarrow \infty }\int \dd^3x
&  \Bigg[ \delta \psi_1   \left(\frac{1}{2} \beta  f \delta g_{rx}-\frac{1}{2} f r^2 \delta\psi_1'\right)+ \frac{2}{r} \delta g_{tx}^2 -\frac{1}{2} f  {\delta A}_x {\delta A}_x'    \\
& - \delta g_{{tx}}   \left(\frac{1}{2}  \delta\dot{g}_{{rx}}-\frac{1}{2} r^2 (\frac{\delta g_{tx}}{r^2})' +\frac{\mu}{2 r^2}    \delta A_x\right)\\
& +\eta _c \left(\delta \psi_1  \left(\frac{r^2 \ddot{\delta \psi_1 }}{2 \sqrt{f}}-\frac{\beta   \dot{\delta g}_{tx}}{2 \sqrt{f}}\right)+\frac{\beta   \dot{\delta \psi_1 } \delta g_{tx}}{2 \sqrt{f}}-\left(\frac{2 }{\sqrt{f}} \right)\delta g_{{tx}}^2 \right)  \Bigg] ,
\end{split}
\end{equation}

where $f(r)= r^2 -\frac{\beta^2}{2} -\frac{m_0}{r}+ \frac{\mu^2}{4r^2} $.  We dropped the boundary contribution from the horizon as a prescription for the retarded Green function \cite{Son:2002sd}.\footnote{In fact, the contribution of the incoming solution at the horizon is zero in \eqref{Sreal}, which is real. However, for a generating function of {\it retarded} Green's functions, we will take only part of \eqref{Sreal} as explained below \eqref{action1}, which is complex.  In this case, it turns out that the contribution from the horizon is pure imaginary. From this perspective,
we should drop the contribution from the horizon.  }
Near boundary $r \rightarrow \infty$, the fluctuation fields in momentum space, \eqref{fluc1} - \eqref{fluc4},  may be expanded as
\begin{equation}
\begin{split}
&  h_{tx}(\omega,r) = \sum_{n=0}^{\infty} \frac{ h_{tx}^{(n)}(\omega)}{r^n}  \,, \qquad
    h_{rx}(\omega,r)=  \sum_{n=0}^{\infty} \frac{  h_{rx}^{(n)}(\omega)}{r^n} \,, \\
&   { a}_{x}(\omega,r) = \sum_{n=0}^{\infty} \frac{  {  a}_{x}^{(n)}(\omega)}{r^n} \,, \qquad
    \chi(\omega,r) =  \sum_{n=0}^{\infty} \frac{  \chi^{(n)}(\omega)}{r^n}   \,,
\end{split}
\end{equation}
and using the equations of motion, we can obtain a quadratic action as follows
\begin{equation} \label{action1}
\begin{split}
S_{\mathrm{ren}}^{(2)} = \frac{V_2}{2} & \int_0^\infty   \frac{ \dd \omega}{2\pi}
\Bigg[ -\mu  {\bar a_x}^{(0)} h_{tx}^{(0)} -\mu  {\bar h_{tx}}^{(0)} a_{x}^{(0)}   -2 m_0{\bar h_{tx}}^{(0)} h_{tx}^{(0)} + \bar a_x^{(0)} a_x^{(1)}  \\
& + \left(\bar{\chi}^{(0)}  + \frac{i \beta}{\omega} \bar{h}^{(0)}_{tx} \right) \left(3\chi^{(3)} + \beta h_{rx}^{(4)}\right)\\
& +(\eta_c-1) \bigg\{- \Lambda^3 \left(4 {\bar h_{tx}}^{(0)} h_{tx}^{(0)}\right) - \Lambda^2  \left(4{\bar h_{tx}}^{(1)} h_{tx}^{(0)}+4 i {\bar h_{tx}}^{(0)} h_{rx}^{(2)} \omega \right) \\
&  \qquad \ + \Lambda  \left(  i \beta \bar h_{tx}^{(0)}  \chi^{(0)} \omega - 2 i {\bar h_{tx}}^{(0)} h_{rx}^{(3)} \omega + \beta^2 {\bar h_{tx}}^{(0)} h_{tx}^{(0)}\right)  \\
& \qquad \ +\Lambda \left(-4 i {\bar h_{tx}}^{(1)} h_{rx}^{(2)} \omega -4 {\bar h_{tx}}^{(2)} h_{tx}^{(0)}+i \beta  {\bar\chi}^{(0)} h_{tx}^{(0)} \omega -{\bar\chi}^{(0)} \chi^{(0)} \omega ^2\right)\\
& \qquad \ - 2m_0 \bar h_{tx}^{(0)}  h_{tx}^{(0)} - 4 \bar h_{tx}^{(0)} h_{tx}^{(3)} - 2i \omega \bar h_{tx}^{(1)}h_{rx}^{(3)} +\beta^2 \bar h_{tx}^{(1)} h_{tx}^{(0)}+i\beta \omega \bar h_{tx}^{(1)}\chi^{(0)} \\
 & \qquad \  -4 i \omega \bar h_{tx}^{(2)} h_{rx}^{(2)} - 4 \bar h_{tx}^{(3)}h_{tx}^{(0)} + i \beta\omega \bar \chi^{(1)} h_{tx}^{(0)} -\omega^2 \bar \chi^{(1)}\chi^{(0)}  \bigg\} \Bigg] + \Bigg[\text{c.c}\Bigg],
\end{split}
\end{equation}
where the argument of the fields\footnote{ ${\bar a_x}^{(0)}(\omega) \equiv a_{x}^{(0)}(-\omega) = {a^{(0)}_{x}(\omega)}^* $ by the reality condition of $\delta A_x$. The same notation and reality condition apply to all the other fields. } is $\omega$. $V_2$ denotes volume in $x$-$y$ space and [c.c] means the complex conjugated terms. From here, we will drop the [c.c] term since we want to compute {\it retarded} Green's functions~\cite{Son:2002sd}.

The second line is proportional to a gauge invariant combination under \eqref{residual1}.
Furthermore,  one of  the equation of motion including $h_{rx}^{(4)} $ is
\begin{equation} \label{hrx4}
h_{rx}^{(4)} - \frac{1}{\beta^2-\omega^2}\left( 3 i \omega h_{tx}^{(3)} - i\mu\omega a_x^{(0)} - 3\beta \chi^{(3)}   \right) = 0.
\end{equation}
 One can show that \eqref{hrx4} is equivalent to a Ward identity
\begin{equation} \label{ward1}
\nabla_\mu \langle T^{\mu\nu} \rangle + {F_\lambda}^\nu \langle J^\lambda \rangle - \langle \mathcal O^I \rangle \partial^\nu \psi_I =0 \,,\end{equation}
%
%
%
by using  the boundary metric and the other fields in the linear approximation given as follows:
\begin{align}
&\dd s^2 =\eta_{\mu\nu} \dd x^\mu \dd x^\nu +2 h_{tx}^{(0)} e^{-i \omega t}\dd t \dd x~~,~~\left<T^{\mu\nu}\right>=\left<T^{(0)\mu\nu}\right>+ \left<T^{(1)\mu\nu}\right> \nonumber \\
 &F =-i\omega  a_x^{(0)} e^{-i\omega t} \dd t \wedge \dd x~~~,~~~\left< J^\mu\right> =\left< J^{(0)\mu}\right> +\left< J^{(1)\mu}\right> =\left(\mu,0,0\right)+ \left( 0,  a_x^{(1)}  -\mu h_{tx}^{(0)} ~ , 0 \right) e^{-i\omega t}\nonumber\\
 &\psi_I = \left( \beta x, \beta y\right)~~,~~\left<  \mathcal O^I\right> = \left<  \mathcal O^{(1)I}\right> =\left(  3\chi^{(3)} + \beta h_{rx}^{(4)}   ,0\right)e^{-i\omega t}~~,
 \end{align}
where
\begin{align}
\left< T^{(0)\mu\nu} \right>=m_0\left(
 \begin{array}{ccc}
 2&0&0\\
 0&1&0\\
 0&0&1\\
\end{array}
 \right)~,~
 \left< T^{(1)\mu\nu} \right>=\left(  - 2m_0 h_{tx}^{(0)} - 3h_{tx}^{(3)} + i\omega h_{rx}^{(4)}  \right) \left(
 \begin{array}{ccc}
 0&1&0\\
 1&0&0\\
 0&0&0\\
\end{array}
 \right)e^{-i\omega t}~~.
\end{align}
One may ask why Ward identity of the boundary theory is included in the bulk equation of motion.
It is not accidental: The translation, $x \to x+\xi_{0}$  at the boundary theory is imbedded into the bulk diffeomorphism $x \to x+\xi(x)$, which induces the field transformation $\Phi\to \Phi+\delta_{\xi} \Phi$, which in turn is a special case of general variation,  $\Phi\to \Phi+\delta \Phi$.
Now the equation of motion is coming from the  invariance of bulk action $\delta S_{B}=0$ under the general variation,
while the Ward identity is the requirement of the boundary action under the translation $\delta_{\xi_{0}} S_{b}=0$. Because AdS/CFT request $ S_{B}=S_{b}$ at the onshell, the latter is contained in the huge tower of
equation of motion as a tiny piece.

The terms proportional to $(\eta_c-1)$ in (\ref{action1}) include the divergent terms with $\Lambda$, a regularisation parameter, and finite terms without $\Lambda$. A remarkable fact is that  with  the counter term of weight $\eta_{c}=1$,
not only the divergent terms are cancelled, but also all the  $h_{rx}$  dependent finite terms disappears from the on-shell action, as we claimed in the beginning of this section.

\section{Gauge invariance under the residual gauge transformation} \label{sec5}

Our starting point is the action 
\footnote{It  comes from \eqref{Sreal} before we get eq. (\ref{action1}), for which we have to use the equations of motion.}
\begin{align} \label{action3}
S_{\text{ren}}^{(2)}
=\frac{V_2}{2}
\int_0^\infty \frac{\dd \omega}{2\pi}
\bigg[&  -\mu  {\bar a_x}^{(0)} h_{tx}^{(0)}     -2 m_0{\bar h_{tx}}^{(0)} h_{tx}^{(0)}    + {\bar a_x}^{(0)} a_x^{(1)}- 3{\bar h}^{(0)}_{tx} h_{tx}^{(3)} + 3 \bar{\chi}^{(0)}\chi^{(3)}\nonumber\\& + \left( \beta \bar{\chi}^{(0)}  + i\omega \bar{h}^{(0)}_{tx} \right) h_{rx}^{(4)}  \bigg] +c.c ,
\end{align}
 which is   still dependent on residual gauge \eqref{residual1} even after we set $h_{rx}=0$.
Since it is just a constant  shift of the solution $\Phi$,  its effects are only   shifts of zero-th modes and  $\Phi'(r)$ and all of its  modes, especially $(a_{x}^{(1)},  h^{(3)}_{tx}, \chi^{(3)}):=\Pi^{a}$ are intact.  Notice that  the recurrence relations derived  from equations of motion relate higher modes with the zero-th modes  $J^{a} = (a_{x}^{(0)}, h^{(0)}_{tx},\chi^{(0)})$. However, all dependences of
 higher modes on zeroth modes is through the gauge invariant combination
 $\omega \chi^{(0)} -i\beta h^{(0)}_{tx}$. See, for example, \eqref{gaugecom}.
Thus all higher modes are gauge invariant,  which makes the gauge invariance of the $\Phi'(r)$ intact in spite of the complicated dependence of higher modes on the zeroth modes.

 The residual gauge dependence of  \eqref{action3} can be understood as follows.
The full on shell action should be invariant under the residual gauge transformation.
However, what we are looking at is the quadratic part of the action $S^{(2)}_{\text{ren}}$, which generates the 2-point function,  in the expansion of
\be
 S_{\text{ren}}[\delta \Phi]=S^{(0)}_{\text{ren}} +S^{(1)}_{\text{ren}}[\delta \Phi]+S^{(2)}_{\text{ren}}[\delta \Phi] + \cdots,
\ee
where $\delta \Phi=(\delta\Phi_{\mu\nu},\delta \Phi_\mu, \delta\Phi_I)$ collectively denotes the sources of the dual field theory, which are boundary values of $\frac{1}{r^2}\delta g_{\mu\nu}$, $\delta A_\mu$ and $\delta \psi_I$. $S_{\text{ren}}^{(1)}[\delta \Phi]$ and $S_{\text{ren}}^{(2)}[\delta \Phi]$ are given as follows:
\begin{align}
&S_{\text{ren}}^{(1)}[\delta \Phi] = \int \dd^3 x
 \left(  \frac{1}{2} \delta \Phi_{\mu\nu} \left< T^{(0)\mu\nu} \right> + \delta \Phi_\mu \left< J^{(0)\mu} \right> + \delta \Phi_I \left< {\mathcal O}^{(0) I} \right> \right) \,, \\
&S_{\text{ren}}^{(2)}[\delta \Phi] = \int \dd^3 x   \left( \frac{1}{2}   \delta \Phi_{\mu\nu} \left< T^{(1)\mu\nu} \right> + \delta \Phi_\mu \left< J^{(1)\mu} \right>+ \delta \Phi_I\left< \mathcal O^{(1) I} \right>  \right) \,.
\end{align}
Under the residual gauge transformation\footnote{ This transformation changes the sources of the action, $\delta \Phi_{\mu\nu},\delta \Phi_\mu, \delta \Phi_I$ . One should note that there are non-vanishing transformations for $\delta \Phi_{00} $ and $\delta \Phi_0$.} with $h_0 = -i \omega\zeta_0$, the variations of these actions are
\begin{align}
&\delta S_{\text{ren}}^{(1)}[\delta \Phi] = V_2 \int \frac{\dd \omega}{2\pi} \left\{ \bar\zeta_0   \left( i\omega \mu a_x^{(0)} + 2 i \omega m_0 h_{tx}^{(0)} \right)+ c.c \right\}~~,\\
&\delta S_{\text{ren}}^{(2)}[\delta \Phi] = -\delta S_{\text{ren}}^{(1)}[\delta \Phi] + V_2 \int \frac{\dd \omega}{2\pi} \left\{ \bar\zeta_0   \left(3\beta \chi^{(3)}  -3 i \omega h_{tx}^{(3)} +i\omega \mu a_x^{(0)} +\left(\beta^2 - \omega^2\right) h_{rx}^{(4)} \right)+ c.c \right\}~~.
\end{align}
 Thus the total variation is proportional to the Ward identity  \eqref{hrx4}.
Notice that  $S_{\text{ren}}$ is gauge invariant but $S_{\text{ren}}^{(2)}$, which is starting point to derive the
Green function,   is not invariant by itself.
Nevertheless   physical observables derived from $S_{\text{ren}}^{(2)}$ are invariant because
the Green functions are second derivatives of the full on shell action at the zero source limit.

At this point one can discuss   a puzzle in counting degrees of freedom. There are only two independent bulk solutions satisfying the in-falling boundary conditions,\footnote{We have two second order differential equations and one first order one in three variables: $a_x, h_{tx}, \chi$.
Therefore, there are 5 boundary conditions to fix.
If we fix the in-falling boundary conditions  for all three variables, we are left with two degrees of freedom.
We recall the equations \eqref{nearh} and \eqref{const1}.  }
 while we need three  solutions at the boundary
 since there are three independent  source fields.
 Therefore, there is a crisis of mismatch of degrees of freedom between the bulk and boundary. What solves the problem is the RGS \eqref{RGS}.
We call it RGS because it is generated by the zero mode of a diffeomorphism generator.
On the other hand, to be a proper gauge degree of freedom in the bulk, the  diffeomorphism generator should satisfy the proper boundary conditions: infalling at horizon and Dirichlett at boundary.
The residual gauge symmetry generator is a global shift and therefore it can   satisfy neither of them.
So such a shift by the diffeomorphism zero mode is not a true gauge symmetry, while  it is a symmetry of the bulk equations of motion.
In other words,  the RGS    is a ``solution generating symmetry"   rather than a gauge symmetry.
  Therefore, the gauge orbit of RGS {\it can} provide us
   the necessary degree of freedom (d.o.f) near boundary. To match the d.o.f, we need to accept its bulk orbit as
 physical configuration inspite of the fact that the resulting bulk solution does not satisfy the infalling BC.
  \footnote{
 So far we  discussed the degree of freedom mismatch using the RGS, since our formalism in \cite{Kim:2014bza} to calculate the conductivity  is based on it.}
 One can give a more natural bulk solution
 by extending RGS to a diffeomorphism  which satisfies the infalling boundary condition and it is reduced to our previous RGS near the boundary.
It is generated by $\xi^\mu = (0,  \zeta(r)e^{-i\omega t},0,0)$, with\footnote{ We thank the anonymous referee for suggesting to consider this.}
\be
  \zeta(r)=\epsilon (f(r)/r^2)^{-i\omega / (4\pi T)}, \label{BSD} \ee
  where $f$ is the metric factor given in Eq. (\ref{groundsol2}) and $\epsilon$ is a constant parameter.
 Notice that the RGS is the case where $\zeta(r)$ is constant.
We will call this ``boundary shifting diffeomorphism''(BSD).
Now we can understand the degree of freedom mismatch as follows:
Since it is not satisfying the Dirichlett bc, it is still not a proper gauge transformation.
Notice also that under (\ref{BSD}), the gauge slice is shifted and some of the gauge fields become
singular. For the discussion on the treating these issues, we refer the reader to page 24 of ref. \cite{Donos:2014uba}.
\footnote{It is very tempting to consider BSD as a gauge transformation at least from bulk point of view.
If we do it, we get to the problem:  Its orbit in the boundary generate physical configuration while
it does not in the bulk, so that crisis of d.o.f becomes real!}
This is   the reason why the BSD can generate a new solution in the boundary.
It is precisely the same logic why RGS generate new solution. \footnote{ { This argument is  further justified if we consider the numerical calculation starting from the boundary instead from horizon.
After choosing 3   fields's values, we can adjust two ``expectation values'' such that we can get
infalling boundary values at the horizon.   It is easy to show that only when we start from a subspace of codimension 1, we get three infalling solution near the horizon. If we start from a point off this plane,
we get one infalling and two fields which are mixture of infalling and a constant. In this calculation the gauge condition  $h_{rx}=0$ is intact.
This demonstrates that we can not impose infalling bc for all fields at hands.
If we do the same numerical experiment for BSD, the picture is following. The BSD generate the the orbit
and it also move the gauge slice.  Now in this case even in the case we start from the off the plane,
we can get three infalling fields at the horizon.
We need to calculate the r-evolution at each   `gauge fixing' plane which pass through the initial data.
 }}
 Since RGS and BSD shift the boundary values of fields, they generate the Ward identity for  the translation invariance.
This  is a typical example how a global symmetry is encoded in a local gauge transformation and  how the apparent paradox of the degree of freedom  can be resolved because of the holographic correspondence.
\footnote{The apparent `mismatch' is due to the difference in viewing the gauge orbit of BSD (or RGS) between the
 bulk and boundary. In the bulk, one could view  it as  gauge orbit.
 On the other hand, from the boundary theory point of view, there is no gauge structure and
 the orbit of translation symmetry is physical degree of freedom. }

\section{Basis independence}
In \cite{Kim:2014bza}, we constructed a formalism to perform the AC conductivities for the case where multiple fields are coupled together. We had to choose a basis of initial conditions and one can ask whether different choices
of basis give the same result.  Answering this question will also provide an alternative reasoning of gauge invariance.
 To provide the setup, let us consider $N$ fields  $\Phi^a(x,r)$, ($a=1,2,\cdots, N$),
\begin{equation}
\Phi^a(x,r) = \int \frac{\dd^d k}{(2\pi)^d}  e^{-ikx}  r^p \Phi^a(k,r)\,, \label{newphi}
\end{equation}
where the index $a$ may include components of higher spin fields. For convenience, $r^p$ is multiplied such that the solution  $\Phi^a(k,r) $ goes to constant at boundary. In our case, $(\Phi^1, \Phi^2,\Phi^3)  = (a_x, h_{tx},  \chi)$ and $p=0$ for $\Phi^1, \Phi^3$ and $p=2$ for $\Phi^2$.

Near horizon($r=1$), solutions can be expanded as
\begin{equation} \label{incoming}
\Phi^a_i(k,r) = (r-1)^{\nu_{a\pm}} \left( \varphi^{a}_i + \tilde{\varphi}^{a}_i (r-1) + \cdots \right) \,,
\end{equation}
where  a new subscript $i$ is introduced to denote the solutions corresponding to a specific independent set of initial conditions.
For example, $\varphi^{a}_{i}$  may be chosen as
\begin{equation} \label{init}
\begin{split}
\varphi^{a}_{1}&= \left(1, \ -({\tilde \mu} +  {\tilde \beta} )/(1+\nu),\ \ \ 1\right)^\intercal \,, \\
\varphi^{a}_{2}&=\left(1, \ -({\tilde \mu} - {\tilde \beta} )/(1+\nu) ,\ -1 \right)^\intercal \,,
\end{split}
 \end{equation}
 where  we used \eqref{const1} and $\nu =  - i\omega/(4\pi T)$ as shown below \eqref{nearh} for incoming boundary condition to compute the retarded Green's function \cite{Son:2002sd}. Due to incoming boundary condition, $\varphi^a_i$ determines $\tilde{\varphi}^a_i$ through horizon-regularity condition so that we can determine the solution completely.
 Each  initial value vector  $\vec{\varphi}_i$   yields a   solution, denoted by $\vec{\Phi}_i(r)$, which is expanded as
\begin{equation} \label{sando}
\Phi_i^a(k,r)  \rightarrow   \mathbb{S}_{i}^{a}  + \cdots +  \frac{\mathbb{O}_{i}^{a}}{r^{\delta_a}}  + \cdots   \qquad (\mathrm{near\ boundary})\,,
\end{equation}
where  $\mathbb{S}_i^a$ are the {\it{sources}}(leading terms) of $i$-th  solution  and  $\mathbb{O}_{i}^{a}$ are the {\it{operator }}expectation values corresponding to sources($\delta_a \ge 1$).

  Notice that we have only two solutions while we have a three dimensional vector space  $\cal J$
  of boundary values $J^{a}$, $a=1,2,3$.
  To fix such mismatch of degree of freedom, we introduce a constant solution
  $\vec{\Phi}_{0}(r)= \vec{\mathbb{S}}_{0}=(0,1,i\beta/\omega)^\intercal$ along the gauge-orbit direction of the residual gauge transformation so that
$ \mathbb{S}_{1}^{a},  \mathbb{S}_{2}^{a}, \mathbb{S}_{0}^{a}$ form a basis of $\cal J$.  Now $\mathbb{S}$ and $\mathbb{O}$  are generic regular matrices of order 3.

The general solution is a linear combination of them: let
\be \Phi^a(k,r) = \Phi_{i}^{a}(k,r) c^i \,, \ee
 with  real constants $c^i$'s. We can choose $c^i$ such that the combined source term matches the boundary value $J^a$:
  \begin{equation} \label{source}
J^a = \mathbb{S}_{i}^{a} c^i \,,
\end{equation}
 which yields
\begin{equation} \label{GS}
\Phi^a(k,r) = \Phi_{i}^{a}(k,r) c^i
\rightarrow J^a + \cdots + \frac{\Pi^a}{r^{\delta_a}} + \cdots \,,   \qquad (\mathrm{near\ boundary})
\end{equation}
where, with \eqref{sando} and \eqref{source},
\begin{equation} \label{Pi1}
\Pi^{a}= \mathbb{O}_{i}^{a} c^i = \mathbb{O}_{i}^{a} (\mathbb{S}^{-1})^{i}_{b} J^b =: C^a_b J^b \,.
\end{equation}
Notice that both $\Pi^a$  and $C^a_b$ are invariant under  the transformation $ {J}^b \to J^{b}+\epsilon S_{0}^{b} $ because   $C^a_b S_{0}^{b}  =  \mathbb{O}_{i}^{a} (\mathbb{S}^{-1})^{i}_{b}S_{0}^{b} =   \mathbb{O}_{0}^{a} = 0$, where $\mathbb{O}_{0}^{a} = 0$ since it is the sub-leading term of the constant solutions.

A general on-shell quadratic action in momentum space has the form of
\begin{equation} \label{sb}
S_{\mathrm{ren}}^{(2)}
= \frac{1}{2} \int \frac{\dd^d k}{(2\pi)^d}  \left[ \bar{J}^a \mathbb{A}_{a b}(k) J^b
+  \bar{J}^a \mathbb{B}_{a b}(k) {\Pi^b} \right]  ,
\end{equation}
where $\mathbb{A}$ and $\mathbb{B}$ are regular matrices of order $N$.  $\bar{J}^a$ means $J^a(-k)$ and,
in matrix notation, $\bar{J}^a$ can be understood as a row matrix.
For example, in our case,  the effective action \eqref{action3} reads
\begin{equation} \label{517}
S_{\mathrm{ren}}^{(2)} = \frac{V_2}{2}\int \frac{\dd \omega}{2\pi}  \left[ \bar{J}^a \mathbb{A}_{a b}(\omega) J^b
+  \bar{J}^a \mathbb{B}_{a b}(\omega) {\Pi^b}  \right],
\end{equation}
where
\begin{align}
J^a =
\begin{pmatrix}
    a_x^{(0)}  \\
    h_{tx}^{(0)} \\
   \chi^{(0)} \\
\end{pmatrix}\,, \quad
\Pi^a =
\begin{pmatrix}
    a_x^{(1)}  \\
    h_{tx}^{(3)} \\
   \chi^{(3)} \\
\end{pmatrix}\,, \quad
\mathbb{A} = \begin{pmatrix}
    0 & -\mu& 0  \\
    -\mu & -2 m_0& 0  \\
   0 & 0& 0  \\
\end{pmatrix}\,, \quad
   \mathbb{B}= \begin{pmatrix}
 1 & 0 & 0 \\
 0 & 0 & \frac{3 i \beta}{\omega} \\
 0 & 0 & 3 \\
\end{pmatrix} \,.
\end{align}

With \eqref{Pi1}  the action  \eqref{sb}   becomes
\begin{equation} \label{Gab}
\begin{split}
S_{\mathrm{ren}}^{(2)}
 &= \frac{1}{2}  \int_{\omega \ge 0} \frac{\dd^d k}{(2\pi)^d}  \left[ \bar{J}^a \left[\mathbb{A}_{a b}(k) + \mathbb{B}_{ac}\mathbb{O}_{i}^{c} (\mathbb{S}^{-1})^{i}_{b}(k)\right]J^b  \right] \\
& \equiv \frac{1}{2}  \int_{\omega \ge 0} \frac{\dd^d k}{(2\pi)^d}  \left[ \bar{J}^a  G_{ab}^R J^b  \right] \,,
\end{split}
\end{equation}
where the range of $\omega$ is chosen to be positive following the prescription in  \cite{Son:2002sd}.
Notice that  $ \mathbb{O}_{i}^{a} (\mathbb{S}^{-1})^{i}_{b} $ is independent of $J^{a}$, because neither $\mathbb{S}_{i}^{a} $ nor $\mathbb{O}_{i}^{a} $ depends on $J^{a}$.
Furthermore $ \mathbb{O}_{i}^{a} (\mathbb{S}^{-1})^{i}_{b} $ is independent of the choice of the initial condition
 \eqref{init}, because the different choice of initial value vectors are nothing but a linear transformation $ {\varphi^a_i} \to {\varphi^a_j}R^j_i$, which induces right multiplications in the solutions:
$\mathbb{S} \to \mathbb{S} R, \mathbb{O} \to \mathbb{O} R $.
This proves the Green functions are independent of choice of basis for our initial conditions.

Notice that since $\mathbb{A}$ and $\mathbb{B}$ are also independent of $J$, $G_{ab}^R$ is independent of $J$ and manifestly gauge invariant, giving alternative reason for the invariance of the Green functions under the residual gauge symmetry.

\section{Conclusion}

We investigated  the gauge  invariance of physical observables in a holographic theory
under the local diffeomorphism. We find that gauge invariance is closely related to the holographic renormalisation.
Apart from the zero-th mode residual gauge dependence, gauge dependence is cancelled by the local counter terms defined in the boundary,
However, due to the difference in the space-time structure between the near-horizon and near boundary regions,  there are  residual gauge structure near boundary. There is a subtle and deep connection between the degrees of freedom at the boundary and those at the bulk.
There are three degrees of freedom   at the boundary, out of which
  only two  can be embedded into bulk fields such that they are the boundary values of the  bulk fields satisfying the incoming boundary conditions.
The residual gauge symmetry is not  a proper gauge symmetry but a solution generator {\it near} the boundary.
We proved the   invariance of Green's functions under such a symmetry  in the context of  algorithm by which all AC transports are constructed simultaneously.

We can extend the RGS such that it satisfies the infalling boundary condition, which we call the boudanry shifting diffeomorphism.
Then we can view things more concisely and natural.
 RGS is not gauge symmetry but a solution generating transformation. Therefore it generate formally new solution both in boundary and bulk. By extending it to BSD, the bulk part of the solution can  be accepted as a true bulk degree of freedom more naturally since the latter     satisfies the in-falling boundary condition.

\vskip 0.5cm
\noindent
{\bf Note added: }
After this work is almost finished, the paper \cite{Donos:2014gya} appeared where
residual gauge invariance was discussed using a different method.

\section*{Acknowledgments}
The work of KYK and KKK was supported by Basic Science Research Program through the National Research Foundation of Korea(NRF) funded by the Ministry of Science, ICT \& Future Planning(NRF-2014R1A1A1003220).
The work of SS and YS was supported by Mid-career Researcher Program through the National Research Foundation of Korea (NRF) grant No. NRF-2013R1A2A2A05004846 and
SS was  also  partially supported by the Korean-Eastern European cooperation in research and development through  (NRF-2013K1A3A1A39073412).
YS was also supported in part by Basic Science Research Program through NRF grant No. NRF-2012R1A1A2040881.

\providecommand{\href}[2]{#2}\begingroup\raggedright\endgroup


\begin{thebibliography}{10}

\bibitem{Witten:1998qj}
E.~Witten, {\it {Anti-de Sitter space and holography}},  {\em Adv. Theor. Math.
  Phys.} {\bf 2} (1998) 253--291,
  [\href{http://xxx.lanl.gov/abs/hep-th/9802150}{{\tt hep-th/9802150}}].

\bibitem{Hartnoll:2009sz}
S.~A. Hartnoll, {\it {Lectures on holographic methods for condensed matter
  physics}},  {\em Class.Quant.Grav.} {\bf 26} (2009) 224002,
  [\href{http://xxx.lanl.gov/abs/0903.3246}{{\tt arXiv:0903.3246}}].

\bibitem{Kim:2014bza}
K.-Y. Kim, K.~K. Kim, Y.~Seo, and S.-J. Sin, {\it {Coherent/incoherent metal
  transition in a holographic model}},
  \href{http://xxx.lanl.gov/abs/1409.8346}{{\tt arXiv:1409.8346}}.

\bibitem{Kim:2015dna}
K.-Y. Kim, K.~K. Kim, and M.~Park, {\it {A Simple Holographic Superconductor
  with Momentum Relaxation}},  \href{http://xxx.lanl.gov/abs/1501.0044}{{\tt
  arXiv:1501.0044}}.

\bibitem{Amado:2009ts}
I.~Amado, M.~Kaminski, and K.~Landsteiner, {\it {Hydrodynamics of Holographic
  Superconductors}},  {\em JHEP} {\bf 0905} (2009) 021,
  [\href{http://xxx.lanl.gov/abs/0903.2209}{{\tt arXiv:0903.2209}}].

\bibitem{Kaminski:2009dh}
M.~Kaminski, K.~Landsteiner, J.~Mas, J.~P. Shock, and J.~Tarrio, {\it
  {Holographic Operator Mixing and Quasinormal Modes on the Brane}},  {\em
  JHEP} {\bf 1002} (2010) 021, [\href{http://xxx.lanl.gov/abs/0911.3610}{{\tt
  arXiv:0911.3610}}].

\bibitem{Andrade:2013gsa}
T.~Andrade and B.~Withers, {\it {A simple holographic model of momentum
  relaxation}},  {\em JHEP} {\bf 1405} (2014) 101,
  [\href{http://xxx.lanl.gov/abs/1311.5157}{{\tt arXiv:1311.5157}}].

\bibitem{Donos:2014cya}
A.~Donos and J.~P. Gauntlett, {\it {Thermoelectric DC conductivities from black
  hole horizons}},  \href{http://xxx.lanl.gov/abs/1406.4742}{{\tt
  arXiv:1406.4742}}.

\bibitem{Donos:2014uba}
A.~Donos and J.~P. Gauntlett, {\it {Novel metals and insulators from
  holography}},  {\em JHEP} {\bf 1406} (2014) 007,
  [\href{http://xxx.lanl.gov/abs/1401.5077}{{\tt arXiv:1401.5077}}].

\bibitem{Bianchi:2001kw}
M.~Bianchi, D.~Z. Freedman, and K.~Skenderis, {\it {Holographic
  renormalization}},  {\em Nucl.Phys.} {\bf B631} (2002) 159--194,
  [\href{http://xxx.lanl.gov/abs/hep-th/0112119}{{\tt hep-th/0112119}}].

\bibitem{Donos:2013eha}
A.~Donos and J.~P. Gauntlett, {\it {Holographic Q-lattices}},  {\em JHEP} {\bf
  1404} (2014) 040, [\href{http://xxx.lanl.gov/abs/1311.3292}{{\tt
  arXiv:1311.3292}}].

\bibitem{Hartnoll:2007ai}
S.~A. Hartnoll and P.~Kovtun, {\it {Hall conductivity from dyonic black
  holes}},  {\em Phys.Rev.} {\bf D76} (2007) 066001,
  [\href{http://xxx.lanl.gov/abs/0704.1160}{{\tt arXiv:0704.1160}}].

\bibitem{Son:2002sd}
D.~T. Son and A.~O. Starinets, {\it {Minkowski space correlators in AdS / CFT
  correspondence: Recipe and applications}},  {\em JHEP} {\bf 0209} (2002) 042,
  [\href{http://xxx.lanl.gov/abs/hep-th/0205051}{{\tt hep-th/0205051}}].

\bibitem{Donos:2014gya}
A.~Donos, J.~P. Gauntlett, and C.~Pantelidou, {\it {Conformal field theories in
  $d=4$ with a helical twist}},  \href{http://xxx.lanl.gov/abs/1412.3446}{{\tt
  arXiv:1412.3446}}.

\end{thebibliography}
\end{document}